# Attempt to Salvage Multi-million Dollars of Ill-conceived HPC System Investment by Creating Academic Cloud Computing Infrastructure.
# A Tale of Errors and Belated Learning.

Marek Michalewicz[1]

***Abstract*** — In 2015 the Interdisciplinary Centre for Mathematical and Computational Modelling (ICM), University of Warsaw built a modern datacenter and installed three substantial HPC systems as part of a 168 M PLN (36 M Euro) OCEAN project. Some of the systems were ill-conceived, badly architected and for the five years of their life span have brought minimal ROI. This paper reports on a two-year intensive effort to reengineer two of these HPC systems into a hybrid, multi-cloud solution called A-CHOICeM (Akademicka CHmura Obliczeniowa ICM[2]).

The intention was to expand the user base of ICM's typical HPC system from around 200-500 to about 100,000 potential general academic users from all institutes of higher learning in the Warsaw area.

The main characteristics of this solution are integration of on-premises ICM Cloud with several public cloud providers, building solution tailored to particular groups of academic users, containerization, integration of special computational paradigms like AI and Quantum Computing. Full process of designing the solution, competitive dialogue with suppliers, and *full final specifications for the solution* are presented. Several roadblocks, pitfalls and difficulties encountered along the way – including the conservative attitude of "the old school" HPC admins, University bureaucracy, national funding policies and others are presented.

## I. Introduction

AT the end of 2015 the Interdisciplinary Centre for Mathematical and Computational Modelling (ICM), University of Warsaw completed a modern 10,000 sq. m. of technical space datacenter (Fig.1) and installed three substantial HPC systems as part of a 168 M PLN (36 M Euro) OCEAN project. The systems were: Cray XC40 (Okeanos) supercomputer, Huawei cluster (Enigma) specifically targeted for storage and analysis of Big Data using Apache Spark and Hadoop. Enigma consisted of about 350 dual socket, 12 cores CPUs, servers, each with 24 TB local disk space with 43 TB total RAM and 8 PB disk storage (Fig.2). The other cluster "Topola", also from Huawei was meant to serve HPC loads. It consisted of 240 servers with 6 720 cores, 23 TB RAM and nominal computational performance of almost 300 TFLOPs.

Although both Cray Okeanos and Huawei Topola computers for the first three years of their operations (2016–2018) were utilized to their full capacity, the Enigma Huawei cluster had never been utilized in more than about 20% computational and about 12% storage capacity. This situation of wasted investment and unused substantial compute and storage resources was hard to tolerate and to accept by the author who became the ICM Director in 2018. However, any remedy required finding own sources of project funding which were notoriously scarce. An opportunity arose in the beginning of 2020 when ICM started generating own income and there were sufficient funds to start the upgrade project to convert Enigma and some other computer systems together with substantial storage resources into a private cloud service.

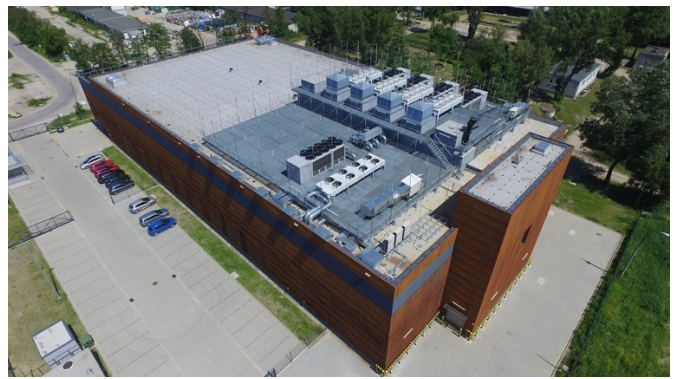

Fig. 1 New ICM datacenter in Warsaw

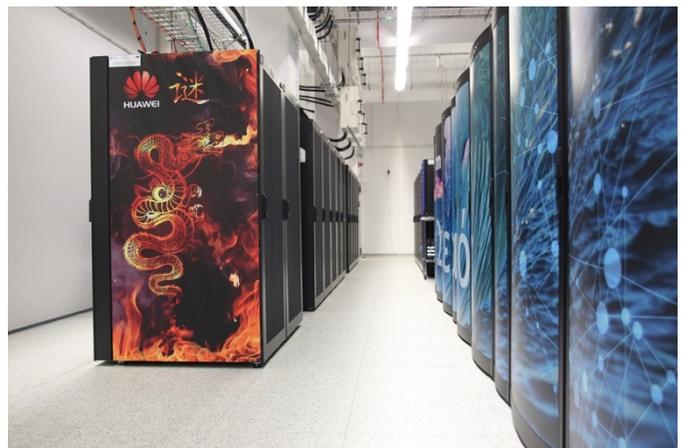

Fig. 2 Enigma BigData system from Huawei (left) and Okeanos Cray XC40 supercomputer at ICM datacenter.

---

[1] Marek Michalewicz was the Director of Interdisciplinary Centre for Mathematical and Computational Modelling, University of Warsaw, between July 2018 and 31st December 2021. Currently he is the Founder and CEO of Siranga Sp. z o.o., Warsaw, Poland, m.michalewicz@siranga.eu

[2] Eng. Academic Computational Cloud of ICM

## II. Cloud Services in Academia and Research

Cloud computing offers many substantial advantages over on-premise computing resources [1-3]. In High Performance Computing world distributed resource sharing was initially realized through Grid Computing paradigm. Especially notable example was the TeraGrid initiative in the US [4,5], which was subsequently transformed into XSEDE initiative and infrastructure [6,7]. These US National Science Foundation's initiatives are currently continued under Advanced Cyberinfrastructure Coordination Ecosystem (ACCES) program [8]. Grid computing paradigm, which in a sense had features of a precursor of cloud computing, was widespread in HPC sector and academic computing centers, for example in Poland it was realized under PL-Grid program [9]. Nimrod/G was an example of a solution enabling resource management and sharing at global scale computational grid [10]. The convergence of Open Data, Open Science and the desirability of sharing of computing resources and computing infrastructure resulted in several national, regional and international initiatives combining Open Data and Cloud Computing paradigms [11-18]

The HPC Cloud Computing concepts, technology and solutions were offered by the notable initiative of UberCloud [19,20] and commercial offerings in a form of portals, UX solutions and resources integrating and provisioning solutions. Rescale [21] was the first such commercial solution followed later by XTREME-D [22,23] and Ronin [24], among others. An important example of an experimental large scale Cloud Computing is Chameleon [25,26] developed at Argonne National Laboratory.

## III. Problems of Purpose and Design

The problem of very low utilization of Enigma, which never exceeded 20% of compute resources and no more than 12% of total disk storage was a result of several serious design and purchase errors: i) the promise of widespread adoption of MapReduce Hadoop anticipated at the time of planning the infrastructure requirements in 2014 never materialized; ii) the design of architecture of the 10–racks, 5–modules large Big Data Enigma system had several critical faults that were almost impossible to eradicate without a serious reengineering of the architecture: first – the system was not multi-user nor multi-tenant since data within one module (two racks or about 70 servers) could be overwritten by another user on the same module; and finally the external data link to the world was only 1 Gbps, meaning that in order to fill the 8PB of local disk storage available for Big Data processing on Enigma, it would take over two years of continuous ingest of data. The InfiniBand interconnect had rather awkward topology and was not able to support fast movement of data from outside nor within the cluster and the Ethernet network was designed for management only and could not carry any data traffic.

It must be noted that one module of Enigma has been used for OPENAire project where ICM played a role in providing hardware resources [1].

## IV. Problems of Internal Resistance and Systems Roadblocks

Shortly after joining ICM as a deputy director in late 2016, the author first became aware of the problem of an idle HPC resources and incredible wastefulness of substantial publicly funded infrastructure. When he begun discussions of this issue with his colleagues, he met with indifference and discomfiture. The technical staff disowned the problem, the decision makers appeared not to be responsible nor concerned. In 2018 when the author was appointed the Acting Director of ICM it was possible to launch a project to re-engineer Enigma purpose and configuration and to integrate some 20+ PB siloed storage into less fragmented storage system. However, it was easier said than done. Many obstacles and problems had to be tackled and some of them were:

i) the conservative attitude of "the old school" HPC admins among ICM technical admin staff. We[3] may list some reasons for this reluctance to adopt changes:
   [START]
   – Cultural Resistance: Many HPC admins are comfortable with the way things have been done traditionally and may resist change, especially if they feel that their expertise is being devalued or replaced.
   – Control: HPC administrators may be accustomed to having complete control over their hardware, software, and data, and may be hesitant to entrust these critical components to a third-party cloud provider.
   – Performance: HPC applications require high levels of performance and low latencies, which may be difficult to achieve in a cloud environment due to the virtualization and network overhead. HPC administrators may be concerned that moving to the cloud will negatively impact performance.
   – Cost: Cloud providers typically charge for compute and bandwidth usage, which can quickly add up for HPC workloads that generate large amounts of data.
   – Complexity: Cloud computing introduces additional layers of complexity to HPC workflows, including networking, security, and data management. HPC administrators may be concerned about the additional time and effort required to manage these new components. [END][4]

Some ICM admin and technical staff shared above objections and we engaged in continuing dialogue to reach consensus and to get all experienced, but sometimes obstinate members of technical team on-board.

Another objection brought by critics of an idea was that this project can be realized internally entirely within our own organization. The author vehemently objected this idea arguing that:

---

[3] The question the author asked CgatGPT was: "Why do "the old school" HPC admins resist cloud computing?

[4] The text between [START] and [END] was written by ChatGPT. I would have provided the same arguments myself.

a. commercial cloud providers are years ahead with their solutions and technology and entities from this environment were the most suitable providers of the solution we were seeking,
b. there was little research and innovation value for ICM in this kind of development project;
c. ICM's technical staff was already intensely involved in their routine duties and multitude of research projects;
d. ICM should not serve as a software development house, but as a research institution and all effort of technical stuff should be directed to support of research, not software development project, and;
e. our staff could engage in back-room "Open Stack", virtualization and similar development but was inexperienced in accounting, user expectations UX, and immediate needs and font end management of a very large group of expected new users. HPC practitioners are aware of occasional admins-users polarization which might not translate well into "user-friendly" products.

ii) University bureaucracy.
Bureaucrats dislike changes and exceptions. The comfortable, steady state with no perturbations and routine operations are the norm – when confronted with unique and special situation, they will become obstructionist and will try to slow or derail the process [27]– that was the case here. It took six months, between June 2020 to end of December 2020, of tedious and occasionally trivial probing by the University of Warsaw Public Procurement Office Manager to allow ICM to open a "Competitive Dialogue". The idea that the procurement does not relate to a "ready-made, out-of-the-box" product but to the original technological solution which has never been implemented in Poland; that the solution provider had to build a consortium of partners to integrate multiple technology components into one solution; and the fact that only very experienced technology players were capable of delivering such a solution was not easy to grasp and accept by the University Procurement Office.

iii) Funding for the project
National funding policies for HPC have changed substantially after Poland joined EuroHPC in 2018. Most funding is channeled through EuroHPC schemes, and the coordination of these activities was delegated by the Ministry of Science and Higher Education[5] to Cyfronet in Krakow [28]. ICM was also ineligible for so called POIR national funding scheme [29] since Warsaw and Mazovian region was deemed developed and some 93% of POIR funding was channeled to under-developed regions. University of Warsaw was also not open to provide extra funds to ICM since the ICM's OCEAN project completed in 2015 accrued cash shortage of the order of 18 M PLN which was treated by the University as the internal debt of ICM. Curiously the incredibly modern and valuable infrastructure built by ICM as part of OCEAN project including the datacenter and all HPC infrastructure was absorbed as the University assets which were never offset against the ICM cash debt.
Starting in 2018 ICM launched several projects which over the next three years generated income sufficiently large to initiate and fund A-CHOICeM project.

iv) Technical incompatibility of components.
The very large local disk storage of 8 PB of Enigma Big Data cluster was difficult to integrate with 12 PB of HPC Lustre storage of Oceanos Cray XC supercomputer. It must be stressed that Tetyda DDN HPC storage system which was part of Oceanos Cray XC40 system was over-specified for ICM users need. Tetyda has 12 PB raw storage capacity out of which not more than 2 PB was ever used until 2021. Within A-CHOICeM project we were looking into ways to utilize this hugely wasted storage capacity for non-HPC academic and research needs. We were also planning to include 240-server Topola into in-house A-CHOICeM cloud infrastructure since most of the workload on this very highly utilized computer were smaller jobs not requiring more than several servers and high percentage of them were request from PL-Grid system which was ripe for cloud-style update anyway.

V. Decision to Convert Big Data and HPC Systems into On-premises Cloud Computing System

The A-CHOICeM project was meant to integrate all underutilized ICM compute and storage resources and to make them accessible to a new and different type of users.
The plan was to provide in-house cloud resources free to all academic users. But a very important feature of the plan was not to stone-wall ICM's private cloud resources from public cloud resources. We insisted on creation of "master portal" which would be accessible to some 100, 000 academic users, initially from the Warsaw region with large number of Universities and Institutes of Higher Learning, which would lead to in-house ICM resources of some 5,000-10,000 virtual machines and ~20PB of storage resources, but also allow to access freely public clouds from all major providers, i.e. AWS, Azure, IBM, Green Lake, etc. Unfortunately, some worthy potential suppliers of a solution were too closely bound to a single public cloud provider and were not able to provide solution open to all public cloud providers.

VI. Competitive Dialogue Process

ICM encountered resistance from the University Public Procurement Office. Essentially the bureaucrats there were not prepared to accept a project of this complexity and novelty. They insisted on a simpler but inadequate tender process.
A formal justification of the request for competitive dialogue was as follows:

"*The Contracting Authority confirms that the competitive dialogue procedure pursuant to 60a(1) will be carried out in*

---

[5] Currently reorganized and named Ministry of Education and Science

*connection with the fulfilment of the conditions pursuant to Article 55(1)(6)-(9) of PPL:*

*6) the solutions available on the market cannot satisfy, without adaptation, the needs of the contracting authority;*

*The order includes the creation of an unprecedented, extremely complex and comprehensive IT solution involving the integration of components from the following areas:*

1. *security of access to vast information resources,*
2. *federalization of access for a huge number of users from different universities of scientific organizations from all over Poland,*
3. *combining HPC solutions with simultaneous provision of these resources in the Cloud Computing modality*
4. *the combination of Cloud storage and Object storage solutions, with Cloud Computing,*
5. *with the integration of batch queuing systems (in HPC) with interactive access, containerization and virtualization.*

*Although each of the above-mentioned components and technologies exists separately on the market - we are not aware of any case on the Polish or European market where all these requirements were met in one installation, simultaneously.*

*The entire solution is a so-called 'bespoke' one and must be closely aligned to the ICM's diverse computer and supercomputing hardware, consisting of around 600 servers, disk arrays of various technologies, different brands with capacities of around 25PB and various file systems - from object-oriented to the high-performance Lustre system. Network solutions also need to be upgraded, both inside the servers and in the LAN/WAN networks, including the supply of network equipment and adapting everything to the very different needs of users. It should also be noted that the solution is to be based as much as possible on existing hardware, with minimal use of new hardware components. We expect that the interconnects in the two clusters will need to be upgraded. An end-to-end solution of this complexity, addressing the very specific needs of a high performance computing (HPC) centre like ICM, is not available on the market.*

*The project will involve not only the purchase of components, but above all the integration and reprogramming, the creation of specialised services.*

*For example, storage vendors (NetApp, Panasas, IBM, HPE, DDN, Western Digital, etc. etc.) have their own hardware-specific solutions, while ICM has storage hardware from all of these vendors - but this hardware is divided by thick "walls" of storage silos. There are few companies in the world that produce data management software that is so-called "hardware agnostic", i.e. independent of the hardware vendor - which is exactly the kind of integrated solution we want in the storage area.*

*Some examples of innovation of the desired solution are:*
- *There is no solution in Poland for making supercomputing resources available as interactive resources. The only possible access is through queue systems such as SLURM, LSF Platform, PBS Pro, or others.*
- *cloud storage combining a High Performance File system (e.g. Lustre, Spectrum) with an object-based file system and SINGLE cloud computing*
- *combination of HPC resources with the ability to create and run jobs using Dockers/Singularity/other containerization systems*

*No such solution exists in Poland.*

*The project will consist not only in reconstruction, but in adding many different elements, software, merging, so called upgrades - i.e. modernisation - e.g. Topola and Enigma clusters must have an upgraded interconnect (internal network). Both these clusters are manufactured by Huawei, but the interconnects are a product of Mellanox (now Nvidia) - so the solutions must be provided by both Mellanox and Huawei.*

*The innovation of the solution will not be limited to the integration of individual elements.*

*We will require the solution to provide access to simulators of quantum computers. That is, the solution requires the possession of such simulators or cooperation with a company that can provide such simulators. It should be emphasized that there are no such providers in Poland and only a few in Europe.*

*As part of the execution of the contract, the contractor will be required to supply additional hardware components. E.g. interconnect components for the Topola and Enigma clusters will certainly be required (optical fibres, switches, network cards, network software).*

*According to our collective knowledge of the high-end market of High Performance Computing, Storage, Networking (HPC, Networking, Storage, Cloud Solutions, System Management, File Systems, Storage Systems, Containerization, Queuing Systems, etc. etc.) there IS NO SUPPLIER that can deliver all the components specified in our order. We expect that even the strongest players in this market will be forced to form their own consortia consisting of several sub-suppliers and sub-contractors.*

*7) the works, supplies or services include design or innovative solutions;*

*Services include design and innovative solutions, combining functionalities for which no integrated solution currently exists. Due to the innovative nature of the solution, its careful pre-design in close co-operation with contractors based on the components offered by them and taking into account the ICM's hardware is necessary. We emphasise that in the case of the current ICM project the whole solution is characterised by great complexity and a very high scale of difficulty, as the solution has to include a lot of new elements and at the same time it has to fit into the existing infrastructure.*

*8. a contract may not be awarded without prior negotiations because of special circumstances with regard to its nature, its complexity or its legal or financial conditions or because of the risks attaching to the works, supplies or services;*

*The order requires prior negotiations with the contractors for technical reasons: it is necessary to ensure full compatibility of the delivered modules with each other and with the ICM hardware at the level of hardware, software and functionalities ensuring full integration with external systems. The ICM also reserves the right to compose a complete, integrated solution according to its own requirements and needs, defined during the dialogue, and not to accept ready-made, out of the box solutions usually offered by suppliers.*

*9) where the contracting authority cannot describe the subject-matter of the contract with sufficient precision by reference to a particular standard, a European Technical Assessment as referred to in Article 30 the description of the subject-matter of the contract in paragraph 1 point 2(c), a Common Technical Specification as referred to in Article 30 the description of the subject-matter of the contract in paragraph 1 point 2(d), or a technical reference.*

*It is clear that elements of the contract will not be ready-made solutions.*
*Any person reading the eligibility criteria and other documents should be someone who is defined as "a person skilled in the art ".*
*Specialists with extensive knowledge of the topics included in this project are fully capable of assessing the project and can guarantee that the supplier is capable of undertaking such a project.*

*It is not possible to describe the subject matter of the contract in accordance with Article 30(1)(d). 2 letter d. because the order involves the integration of dozens of elements, and each of them is non-standard.*

*Due to the technical and conceptual complexity of the contract, as well as the lack of appropriate standards in many areas, it is currently not possible to describe the subject matter of the contract taking into account the technical and legal provisions referred to in Article 30(1)(2)(d) of the PPL, i.e. common technical specifications, understood as technical specifications in the field of ICT products defined in accordance with Art. 13 and Art. 14 reference to provisions of the Civil Code of Regulation (EU) No 1025/2012 of the European Parliament and of the Council of 25 October 2012 on European Standardization, amending Council Directives 89/686/EEC and 93/15/EEC and Directives 94/9/EC, 94/25/EC, 95/16/EC, 97/23/EC, 98/34/EC, 2004/22/EC, 2007/23/EC, 2009/23/EC and 2009/105/EC of the European Parliament and of the Council, and repealing Council Decision 87/95/EEC and Decision No 1673/2006/EC of the European Parliament and of the Council (Official Journal of the European Union L 316 of 14.11.2012, p. 12)*

The selection criteria for Competitive Dialogue partners was created. It is presented in Appendix 1.

The call for Competitive Dialogue was announced in January 2021. The three candidates were selected at the end of January and from early February till July we scheduled weekly one to two-hour sessions with all three potential suppliers. These sessions were followed with internal discussions which led to new sets of questions and problems later in the week communicated to the Dialogue partners. This process lasted till early July 2021.

## VII. Educating Community, Academics and Students and University Officials

ICM was the organiser of four editions of the "Supercomputing Frontiers Europe" series of conferences in years 2018-2021 [30]. The 2021 edition contained a special session on Cloud Computing and HPC with invited talks from UberCloud, Rescale, Ronin and Argonne National Labs [31-35]. ICM has also had organised a series of twenty-one "Virtual ICM Seminars in Computer and Computational Science". As part of the series there was a special lecture delivered by Valerie E. Polichar, Sr. Director & SATO, Academic Technology Services IT Services, University of California San Diego, entitled "Creating a Technology Vision: Planning for the Next Generation of Research Support" [36] which summarised over ten years of experience at UCSD after creation and implementation of the "UCSD Blueprint for the Digital University" [37,38]. ICM was attempting to emulate this process and delivery of the goals defined at that experience at UCSD and to expand computing resources far beyond traditional HPC users.

## VIII. Final Specification of the Solution

The document "DESCRIPTION OF THE SUBJECT MATTER OF THE CONTRACT" is the crux of the present communication and is included in full in Apendix 2.
The A-CHOICeM project has unfortunately never been realised and I believe it is in the public interest of academic centres in Poland and outside to have access to the results of this time-consuming and serious process and its outcome, even if these are only accumulated wisdom and experience, knowledge of the current state-of-the-art in HPC Cloud architecture.

## IX. Conclusion and Lessons Learned

The A-CHOICeM project has never been completed. Our ICM team completed the technical specification document, which is the essential part of the current report, but after resignation of the author from employment at the University of Warsaw at the end of 2021 the project has been discontinued.

During two years of intense engagement in this project and beyond we have seen many of our ideas realized in various projects and solutions – we were following the right path. It is hoped that the presented experiences, specifications and documents will be of some help in other academic and HPC centers and might prove useful even for public cloud providers of academic cloud services and developers of cloud technologies.

APPENDIX 1

TABLE I. SELECTION CRITERIA FOR COMPETITIVE DIALOGUE.

| # | Qualification criterion | Minimum condition | Criterion Weight |
|---|---|---|---|
| 1 | Experience in installation in an HPC environment for academic, scientific or research customers, national laboratories with their own HPC centre (0-10) pts | Implementation of at least two installations in an HPC environment for academic, scientific or research customers such as universities, academic HPC centres, research institutes, national laboratories with their own HPC centre | 10 |
| 2 | Experience in cloud computing installation (0-10 points) | implementation of at least one cloud computing installation with a scale of more than 600 servers | 10 |
| 3 | Experience in installing cloud storage using object storage, at least 10 PB gross (0-10 pts) | implementation of at least one cloud installation for data storage using object storage, at least 10 PB gross | 10 |
| 4 | implementation of at least one multi-cloud installation (including at least two public clouds) | implementation of at least one multi-cloud installation (including at least two public clouds) | 10 |
| 5 | Experience in integrating cloud access with federated authentication (0-10 pts) | implementation of at least one installation with access to clouds with federated authentication system | 10 |
| 6 | Experience in installing cloud computing with integration to the SLURM queuing system (0-10 points) | implementation of at least one cloud computing installation with integration with the SLURM queuing system | 6 |
| 7 | Experience in an installation with an integrated access portal, allowing users to configure virtual machines, containers, SaaS, resources (0-10 pts) | implementation of at least one installation with an integrated access portal allowing users to configure virtual machines, containers, SaaS, resources | 10 |
| 8 | Installation experience with storage data encryption (0-10 points) | implementation of at least one installation with encryption of storage data | 5 |
| 9 | Experience in installation with interface to quantum computing (0-10 points) | implementation of at least one installation with access to quantum computing | 5 |
| 10 | Experience in installation with interface to AI tools (0-10 points) | implementation of at least one installation with an interface to AI tools | 8 |
| 11 | Own and operate a research laboratory in the area of Cloud Computing and HPC | employing not less than 50 persons employed in R&D positions | 6 |
| 12 | Dedicated team (0-10 points) | presentation of at least 4 key persons for the project with a description of their experience | 10 |
| | | | **100** |
| | **Qualification criterion** | **Condition for admission** | |
| 13 | Possession of tools and organisational and technical means necessary for the proper performance of the contract | please present the proposed tools and methodology for project management | – |
| 14 | Possession of certifications for their public cloud services in particular required from such group: SOC 1, SOC 2, SOC 3, CSA STAR, ISO 9001, ISO 27001, ISO 27017, ISO 27018 | please list all the certifications held by the contractor from this list | – |
| 15 | Certificate of clean criminal record of members of the Management Board (current) | | – |
| 16 | Copy of the National Court Register (standard) | | – |
| 17 | Financial report/Profit and loss account for the last 3 years (0-10 points) | Minimum revenue at the level of 80 M PLN | – |

APPENDIX 2

DESCRIPTION OF THE SUBJECT MATTER OF THE CONTRACT

The Order concerns the implementation of cloud computing using the existing ICM infrastructure [hereinafter also referred to as the "Project"] and includes:

a. Implementation of necessary software solutions
b. Supplementation and upgrade of existing ICM equipment
c. Creation of documentation
d. Delivery of training
e. Lifetime support

## DESCRIPTION OF THE SUBJECT MATTER OF THE CONTRACT

*Part 1: Purpose of the project*

*Part 2: Requirements*

1. User access and management portal for system administrators.
2. Security
3. Virtualization, containerization and multi-cloud and hybrid cloud functionality
4. Application sub-portals (areas)
5. Architecture
   i Cloud storage
   ii Network architecture
   iii Cloud architecture and services
   iv Integration
6. Documentation
7. Training
8. Support & Warranty
9. Implementation and payment schedule
10. Way of collecting the order

*(Attachment 1 – The Contracting Authority's Available Infrastructure)*

PART 1: PURPOSE OF THE PROJECT

The intention of this project is to create an easily accessible and as easy to use Academic Cloud Computing ICM (A-CHOICeM) for the academic and scientific community in Warsaw and Poland. The order includes the creation of an unprecedented, extremely complex and comprehensive IT solution involving the integration of components from the following areas:

1. easy, convenient and secure access to vast IT resources,
2. federalization of access for a huge number of users from different universities of scientific organizations from all over Poland,
3. HPC solutions with simultaneous provision of these resources in the Cloud Computing modality (including virtual machines and containerization),
4. Combination of Cloud storage and Object storage solutions with Cloud Computing along with integration of batch queuing systems (in HPC) with interactive access, containerization and virtualization.
5. storage - integrating data storage resources of various types and at the same time making the A-CHOICeM cloud available to users.

Expected features of the A-CHOICeM cloud:

1. Hybridity - The cloud provides equally easy capabilities to create virtual computing resources as well as access to large-scale High Performance Computing resources by integrating the solution with the SLURM queueing system.
2. Multi-Cloud - A-CHOICeM user portal provides access to public clouds such as Azure, AWS, Google Cloud, IBM, GreenLake, etc.
3. Cloud storage - available through a portal, allowing easy creation of storage resources and self-service management of this data - according to specific usage policies.
4. Containerization - the ability to use and create popular containers like Docker and Singularity.
5. Universal and easy access for research workers, teachers and students of Polish universities and research institutes.
6. Access security and security of stored data and user accounts.
7. Well-designed and ready-to-use sub-portals for popular research areas: e.g.
8. quantum chemistry, material sciences, bioinformatics, software development and mathematical modeling (e.g. R, Python, Julia, Mathematica), structural engineering, computational fluid dynamics. Sub-portals must be equipped with, or ready to integrate, the most popular codes, tools, and frameworks in these fields.
9. The A-CHOICeM cloud provides access to state-of-the-art portals/services and tools (both publicly open and, where possible, proprietary) in the fields of Artificial Intelligence (AI) and Quantum Computing.
10. The A-CHOICeM cloud is built using components based on open licenses and allows further expansion and scalability to further local and remote hardware resources and other clouds.

A-CHOICeM is intended to be a solution that will provide access to a much more diverse set of ICM computing resources and allow ICM to open up the entire spectrum of scientific and educational computing and data storage needs of the academic and scientific community to a significantly larger set of users.

This project aims to create a solution that enables easy access and self-management of diverse infrastructure, from a single virtual machine based on single computing cores, to the computing power of supercomputers available at ICM and the diverse resources available in multiple public clouds.

PART 2: REQUIREMENTS

The order for the design and implementation of Academic Computing Cloud ICM – "ACHOICeM" includes the delivery of a technological solution consisting of components that meet the following requirements:

**2.1 User access and management portal for system administrators.**

The primary requirement for the A-CHOICeM access portal must be ease of use and convenience for both users and system administrators.

Federated access must be provided for people from the academic and scientific community; from Warsaw (i.e. UW, PW, WUM, WAT, SGH, SGGW, etc.) as well as from all over Poland.

*Users and usage*
*Cloud users will be:*

1. authorised researchers or students using its computing resources and the software provided therein [**end users**] (with different levels of authorization), and
2. ICM administrators with different levels of authority who manage these resources [**administrators**].

Additionally, depending on the permission level, the administrator will be able to define a catalogue of resources and services available to a given end user or user group. Permissions can be combined into permission groups and assigned to users or user groups.

Depending on the role assigned by the administrator, an end-user may have a different level of authority to share resources or assign roles to other users. For example, a supervisor of a group of users from a university, department, or institute representing that unit will have the authority to share computing and data storage resources assigned to that unit.
In addition, projects, resources and user groups can be defined in the system for a specific period of time. For example, a supervisor nominated by the administrator of a group of students taking part in specific university classes will be able to grant them access to virtual machines and containerized computing environments including specific software, according to appropriately defined access policies.

**Portal functionalities and how to access them**

End-user access to cloud resources will be realized through:

- **web interface** (management of user's cloud resources, access and monitoring of their usage),
- **API**
- **CLI** (ssh) - direct access to user virtual machines allowing authorized users to independently configure virtual machines within the resources allocated by the administrator and monitor the use of those resources.
- **S3 protocol** - access to disk resources

The minimum required user interface features include:

1. user authentication,
2. user profile management,
3. information on the resources allocated and the extent to which they are used,
4. communication with the staff (reporting problems or questions),
5. creation/activation/deletion of virtual machines with user-defined parameters,
6. creating/deleting virtual networks,
7. assigning/releasing virtual machine IP addresses,
8. assigning/releasing volumes for virtual machines,
9. S3 resource creation,
10. creating/activating/deleting containers,
11. Integrated graphical access to disk resources from various file systems (e.g. S3, Lustre),
12. graphical access to HPC systems,
13. Access to integrated public cloud resources through SSO,
14. Access to a code repository (a recipe for building containers),
15. container image repository (general and private for user groups)
16. Access to a service that builds and deposits containers based on prescriptions from the code repository,
17. UI compliance with WCAG 2.0 standard.

The source code and corresponding license of the web interface software must be available to the contracting authority and documented in such a way as to allow for possible future self-modification by the contracting authority.

The web interface must be suitable for use on both mobile devices and desktop computers.

Functionality available to administrators:

1. all end-user interface features,
2. management of users, roles, institutions, projects, mapping, groups/categories/hierarchies of users,
3. software management,
4. management of policies and cloud resources (computing cores, storage, network and limits per user and limits per user group, provisioning period),
5. accounting - accounting for the consumption of resources by users,
6. automatic creation of reports summarizing consumption on the user, group, institution level,
7. monitoring of cloud resources,
8. managing end-user classes, account expiration times, and resource limits,
9. license management.

**2.2. Security**

The ICM cloud infrastructure must meet the following security requirements for the cloud system and stored data:

1. Only authorized users receive access to cloud resources.
2. User authentication is based on SSO mechanism, using 2FA. The implemented solution must allow integration with various external sources of data about users (PIONIER.id, CAS, e-mail addresses in authorized institutions).
   *The 2FA mechanism used must be as universal as possible, available to the greatest possible number of users, without the need to install additional applications or use dedicated hardware devices (tokens) on the user's side (preferred are e.g. SMS one-time codes and/or one-time codes sent to the email address defined in the user's account and confirmed).*
   *At the same time, the 2FA mechanism must be able to cooperate with all implemented mechanisms and sources of user authorization.*
3. Hierarchical separation of access to resources
   *By hierarchical access separation we mean here the ability to configure permissions to system resources and data so that:*
   - *full separation of access to resources for individual tenants was ensured,*
   - *within the resources allocated to a tenant, it was possible to configure separation of access to resources for particular groups of users,*
   - *within resources allocated to groups of users, it was possible to configure separation of access to resources for particular users,*
   - *In both cases (intra-tenant and group), it must be possible to share certain resources.*
4. The services made available by the cloud computing infrastructure on the Internet must be encrypted.
5. The infrastructure should provide event monitoring capabilities for at least the following:
   1. *Availability,*
   2. *use of resources,*
   3. *parameters to help predict the possibility of failure (e.g. S.M.A.R.T),*
   4. *events resulting from software errors (e.g. unexpected interruption of the application).*
6. Central logging of user activity should include at least the following:
   - *enabling/disabling/creating infrastructure elements (virtual machines, containers, virtual networks, assigned disk resources, etc.)*
   - *assigned IP addresses,*
   - *allocated licenses.*
7. The environment must provide the possibility to update the software used (OpenStack, OS, system images for virtual machines, application software, etc.) (see point 11).
8. The infrastructure must provide automatic zeroing of free storage and free memory when a virtual machine is deleted. This should happen completely automatically, without any intervention from users or administrators.
9. The infrastructure must provide backup capabilities at various levels, at least to the following extent:
   - *copies of the entire environment configuration (with rollback capability),*
   - *copies of the portal along with its configuration,*
   - *copies of user data,*
   - *enable backup of the allocated storage on the user / tenant side.*
10. All infrastructure components having a direct impact on business continuity and availability must be redundant.

**2.3 Virtualization, containerization and multi-cloud and hybrid cloud functionality**

1. The portal needs to leverage available resources by accessing virtual machines, containerization, and SLURM managed HPC clusters.

   In terms of virtual machines and containers, the solution will also enable defining the total budget of resources available for use. In particular, the solution must allow defining:
   - *the total number of resource-hours available to the user/group (in particular CPUh/GPUh)*
   - *the period of validity of the budget allocated*
   - *the person responsible for a given pool of resources (project manager)*

   The solution will allow access to the complete billing data (including the budget) through a documented and secure REST API. Access will be possible both to the user's own data (the user himself can check what budget he has and how much resources he used in a given period) and to the full billing data of all users (with appropriate permissions). It must also be possible to fully control available budgets from the same API level by external system (with appropriate permissions). The solution must also have the ability to configure a basic budget available to all users ("free-tier"). The solution must collect and make available data on budget usage with a minimum of hourly resolution. The user must be informed about the approaching end of the budget.

The solution will also provide limit support (per user, per user group), aggregated across all (virtual) machines:

- *the number of simultaneously operating machines*
- *total number of occupied cores*
- *the number of CPU/GPU/Core used by 1 machine*
- *the amount of RAM used*
- *public IP addresses*
- *storage resources (GB, IOPS)*

The billing system will collect data from both virtual machines and containers running on a central containerization cluster.

All configuration items (including e.g. Dockerfile, Ansible scripts or scripts run e.g. in containers or by domain applications) provided by the user should be downloadable/
storage in an integrated private GIT repository, integrated with SSO.

Wherever it is necessary to store VM images or containers, the solution will provide an appropriate mechanism (e.g. an image repository) that will use the S3 resource indicated by the customer as a storage location

2. The portal should realize access to resources and applications in three different variants:

    A. Access to virtual machines

    As part of the portal provided to the users, the end user must be able to run the virtual machine using one of the images provided in the catalog, imported from the public repository or provided on their own. The predefined images in the library must include the current stable release of Ubuntu, CentOS, Rocky Linux, Debian, Alpine and additionally CentOS 7. In addition, the portal will allow the creation of snapshots and copies of the virtual machine. Images provided or imported by the user are included in the total storage taken by the user

    The portal must make available to the user:
    - *SSH access parameters to your own virtual machines*
    - *VNC access through a web browser to own virtual machines (virtual desktop)*
    - *allow parameters to be specified for predefined virtual machines*
    - *allow control of the machines: start, shutdown, reset*
    - *inform (e-mail) the user about upcoming time limits or forced release of resources.*

    B. Access to a container cluster
    The solution will create a centrally managed containerization cluster, allowing users to run images from a local repository, imported from external registries, and build them (by pointing to a public GIT repository or user resource). The local registry (e.g. Harbor) should distinguish images
    provided and supported by the ICM. Images originating from external repositories must be downloaded through the local repository and subjected to automated security checks before being made available to run. Containers running within the central Kubernetes cluster must not originate from sources other than the central registry or external registries via the central registry proxy. The user should be able to choose whether they use their own Kubernetes cluster (on their own budget VMs) or use shared resources to run the container.

    C. Access to an HPC cluster

    Access to resources managed by SLURM Workload Manager within the Okeanos cluster and the Poplar cluster. Within this solution, the selected configuration uses user-provided cluster access parameters. The extent of access and the limits and billing shall be done by an independent system. The delivered solution should support the use of access to the HPC cluster as an element of the pipeline: a job running on the cluster passes the result to the user's virtual machine on which supervised post-processing of the results is performed.

3. Application usage models

    Regardless of the way of access to resources, for each Application defined in the catalogue we distinguish one of two basic usage models: task (batch mode) and service (interactive application). It is possible that a given software is made available in different variants but they are treated as two Applications (e.g. Julia script (task) and Jupyter Notebook).

    Task application/batch mode

    Task-type applications are used to process some predefined task and consume resources until it executes or fails. Predefined task applications are selected by the

user from a directory or are self-created based on generic templates (e.g. Application python script). The user must be able to share his Applications with other users (with a group, with specified users or with the whole Centre). The user configures a set of parameters necessary to run the task, including the indication of disk resources (files, directories) that are to be transferred to the task. The task is then run using the resources defined in the definition of the Application (VM, Container in the central or own environment, HPC). Generic applications that serve as examples and templates for users' own projects must contain examples of Task-type applications running on all available resource types. Task-type applications may be combined by the user into a pipeline, where the output (e.g., directory, tar archive) is passed as input to the next task in the pipeline, defining dependencies, restarting, similar to e.g. GitLab CI/CD Pipelines, Argo Workflows. As an extreme element of such a workflow it should be possible to attach a service application (e.g. virtual desktop for pre/post data processing). As part of the initial installation, Applications must be provided that allow running tasks and analysis using the packages described later in this document.

In the task (batch) mode the user should be able to select the desired application from the category of interest and then upload a set of batch data to the running program. During the task execution the user should be able to control the status of running applications, perform basic operations on them (including cancelling the task), follow the standard output of the program and view the contents of generated output files. The whole process should take place with minimal user's interference, however, the possibility of far-reaching interference should be easily accessible for an advanced user - in particular he should be able to:

- *independently define how the application is invoked by the queue system running on the target node, i.e. directly provide the program name, options and invocation arguments on the command line; • control the standard output, i.e., perform operations on the generated data using a pipeline mechanism, in particular filtering it for desired patterns using the grep tool, the sed stream editor, and other utilities available in the standard environment of POSIX-compliant systems;*
- *request an action to be performed after the application (un)exits correctly, in particular to request an e-mail message to a defined address or to execute a custom script written in a shell language (Bash).*

The task/task mode must also offer the ability to easily (drag & drop) define advanced workflows, including the execution order of several applications and their use of data generated in previous tasks, define conditional instructions based on the outcome of the application, etc.

Example workflow of electron structure calculations - in density functional theory (DFT) - using the Quantum Espresso application. Each of the following steps requires the data generated in the previous step:

- *A self-consistent cycle (SCF) terminated by convergence using the 'pw.x' program - for data and parameters set by the user in a batch file. The system should be able to verify that convergence has been achieved and that it is therefore possible to proceed to the next stage (a conditional instruction written in the shell language - defined by the user).*
- *Non-SCF cycle using the resultant data generated in the SCF cycle. Kohn-Sham states for a given set of 'k' points in the inverse network are computed.*
- *Post-processing of results. When creating a workflow, the user defines which files are post-processed and how: A shell (Bash) instruction or script containing the program call, options, arguments, batch and result data, as well as I/O redirections and pipelines.*

Another typical usage scenario would be to run the application on batch data (uploaded by the user or downloaded from a designated disk resource), and upon completion, run an interactive application running in graphical mode to further process the generated data.

The mechanism for creating and connecting workflow tasks should allow a wide range of configuration options based on the Bash shell language. Between each step it should be possible to define conditional instructions (as well as bigger scripts) and make further execution dependent on them. The goal is a far-reaching automation of work and possibility to execute many dependent tasks after defining them once. The system should allow to create templates of such task pipelines for their later use and modification as well as the use of other popular workflow systems, e.g. Kepler (see also section 5.4: Integration).

4. Service / interactive applications

These types of applications consume resources from startup to user termination. This type of application must be able to share its own resources over the Internet or be restricted to an internal network (shared VPN to the cloud for its users). The basic ways to provide access to the Application are remote desktop access (VNC, WebVNC) and port access e.g. HTTP/S. An interactive Application is run as a VM or as a Container. One Application can run more than 1 VM: e.g. a preconfigured Kubernetes cluster with an internally prepared JupyterHub environment. Depending on the prepared configuration the Application should be able to use the user disk resources or be safely isolated from them. As the end of this Application is related to user action, it is necessary to define for each Application the time of its automatic termination and to be informed about running Applications consuming its resources. The proposed solution must include the following Service/Interactive Applications designed to run on private micro-clusters, automatically configured for the user:

- *Jupyter/Jupyter, Lab/Jupyter Hub, optionally allowing to work with user files. The predefined solution should include a version dedicated to work with students/trainees (sharing after a hash-link with public IP). The application should contain installed Jupyter kernels for languages: Python, Julia and R. It must also allow to choose a container-worker (e.g. with its own Java environment), in which the notebook will be run (standard option of Jupyter Hub),*
- *RStudio, which allows you to work with user files*
- *BinderHub, allowing you to work with public repositories that support it.*

In addition, the solution must include VM images suitable for running the pre/post processing task in fully graphical mode and be based on a current stable release of the Ubuntu and Centos operating systems. Based on the image, the user should have VNC access to the application:

- *ParaView*
- *ANSYS, Matlab and Mathematica packages within the licenses available at ICM or indicated by the user*
- *A clean GNOME Desktop with access to user files and the ability to install software*

The above Applications may be included together as a "Virtual Desktop".

The user should also be able to execute applications running in an emulated text terminal directly from the Portal (without having to set up a standalone SSH session).

5. Scenarios for working with applications

To better illustrate the requirements from the end user's point of view, A-CHOICeM must meet the following usage scenarios:

A. Training using JupyterHub

The lecturer (Registered User) wants to share the interactive Jupyter notebooks with the material prepared for the training by pointing the link to:
- *private/public repository.*
- *private/public docker/singularity image.*

Lecturer can control the amount of shared computing resources per user (number of cores, RAM, workspace capacity, maximum duration of the task / session) and the maximum number of users. The resources consumed by trainees are counted within the lecturer/trainee project (depending on the selected option). Trainees do not need to be A-CHOICeM users, and the lecturer has the option to select "authorization through a known link"

B. Training using a Virtual Machine (VM) image

Lecturer prepared VM image with installed software and batch files necessary to conduct the classes. Trainees have the ability to clone the VM image prepared by the lecturer and run it within their account. Resources consumed by trainees are counted within the project lecturer / trainee (depending on the selected option).

C. Open training

By open training is meant that no user registration is necessary. After providing a magic link (single/per participant) by the lecturer, participants are able to start a JupyterHub or BinderHub session. When creating an event (training), the lecturer sets the "magic link" expiry date.

D. File sharing

Users belonging to the same project can share files via a predefined directory.

E. Status Quo

Users retain the ability to log in via ssh to the HPC cluster and submit jobs to the SLURM queue system, as they did before.

**2.4. Application sub-portals (areas)**

2.4.1 The functionality of the Portal for end users will include running computational tasks within a predefined set of applications. Available applications should be categorized into groups representing research areas. The ability to easily manage categories, applications (add, delete, update) and how to run them should be within the purview of Portal administrators. The key categories and applications that should be among the services offered are listed below (some applications may appear in more than one category).
2.4.2 Availability of an application means the ability to run it in at least one of the access methods described above.
2.4.3 The listing of research areas and applications should be understood as a sample target offering. The portal should implement several typical and welldocumented service delivery patterns, with the possibility to replicate and develop them in the future. In particular, the solution must offer ready-made services to *run the following applications (see also Application Usage Models):*
- Quantum Espresso
- Gromacs
- Mathematica
- Python / Tensorflow / Pytorch
- ANSYS Fluent
- C/C++/Fortran programming environment (GNU, Intel, mathematical libraries)

2.4.4 Target research areas and applications:

*Materials calculations /physics / quantum chemistry*
- *VASP*
- *Wannier*
- *SIESTA*
- *ORCA*
- *NWChem*
- *NAMD*
- *LAMMPS*
- *Gromacs*
- *Gaussian*
- *Quantum Espresso*
- *Elk*
- *Dalton*
- *CP2K*
- *Abinit*
- *CASTEP*

*Molecular dynamics*
- *Gromacs*
- *NAMD*
- *LAMMPS*
- *Charmm*
- *Desmond*

*Computer Algebra / Other*
- *Mathematica*
- *Julia*
- *GNU Octave*
- *Matlab*
- *R*
- *Python*
- *Anaconda*

*CFD: Computational Fluid Dynamics*
- *ANSYS Fluent*
- *OpenFOAM, comes in two versions, both have numerous users:*
  - *The OpenFOAM – [39] (ESI-OpenCFD) latest version v2106.*
  - *The OpenFOAM – [39] (Foundation) - latest v9 and Third Party package.*

*Machine Learning*
- *Tensorflow*
- *PyTorch*

*Programming Languages*
- *Julia*
- *Python*
- *Go*
- *R*
- *Java*
- *GNU: C/C++/Fortran*
- *Intel: C/C++/Fortran*

*Software development, compilers and libraries*
- *GCC: GNU Compiler Collection*
- *Intel (C/C++/Fortran) development toolchain (MKL et al)*
- *BLAS, LAPACK, ScaLAPACK, GSL*
- *OpenMPI, MPICH, MVAPICH*
- *NVIDIA miracles*
- *LLVM (Clang, Flang)*
- *Cmake*
- *Spack*

*Engineering*
- *LS-Dyna*

*Other*
- *Garuda*
- *Taxila*

In addition to the applications available within the existing infrastructure, the Portal should offer the possibility of integration with external service providers in the future, such as

- Wolfram Enterprise Private Cloud - functioning on-premise as part of a private cloud or leading to an external provider-supported service.

- Quantum computing - quantum computers and simulators such as:
  - IBM Q (access to IBM Quantum Network)
  - Quantum Inspire [40]
  - Amazon Braket
  - Forge (D-Wave, Google, IBM)

- AI - Deep Learning (e.g. AWS)

2.5 Architecture

The whole solution belongs to the so-called "made to order" (bespoke) and must be strictly adjusted to the variety of computer and supercomputing equipment owned by the ICM, consisting of approx. 600 servers, disk arrays of various technologies and brands with capacities of approx. 25 PB and various file systems - from object-oriented to high-performance Lustre system.

The proposed solution should take into account the maximum use of existing resources listed in Annex 1 to the OPZ. In the future, this may involve the creation of a multi-stage plan - where the first stage is described in this document, and in the future, other resources may be transformed into the cloud service in subsequent stages of further transformation of the existing resources to the cloud service.

2.5.1 Scale and use of existing ICM infrastructure

ICM Cloud Computing will provide users with virtual computing nodes based on the Customer's existing infrastructure (computing nodes and storage). Cloud should allow for optimal use of all existing equipment of the ordering party, described in Annex 1 to the OPZ.

The system should allow for future expansion with additional resources in the form of additional devices (servers, switches, storage resources).
The implementation documentation should include requirements for hardware that will be compatible with the delivered solution and allow for seamless expansion.

The possibility of future increase of resources (number and type of nodes, disk space) as well as utilization (number of users) to at least 2000 nodes should be foreseen and documented.
In the current stage, the solution should use available Enigma hardware (both for cloud computing and storage) with the possibility of connecting existing Okeanos system resources (via SLURM)

2.5.2 Cloud storage

The Contractor shall implement a data storage resource sharing system ("Cloud storage") based on a single module of the Enigma cluster. The Cloud storage system should be organized in a hierarchical memory structure with the possibility of automatic data migration according to policies defined by administrators. File systems must be integrated and available to individual cloud nodes in accordance with their rights. The system must be fully integrated with other disc resources of the Employer (point 5.4).

1. Technical conditions of the solution
    1. Portal access in accordance with the requirements described in Section 1.
2. Based on object-oriented data storage technologies, compatible with Amazon S3 API
3. Data redundancy and security:
    - choice of replication or erasure coding by object size
    - support for different erasure coding profiles for different objects
    - Automatic or on-demand cluster rebalance, e.g. after adding new disks or removing disks
    - adding and removing disks and servers in a storag cluster without disrupting data sharing
    - uninterruptible upgrades during cluster operation (rolling upgrades)
    - Ability to create policies for stored data (life cycle policy, WORM, etc.)
4. Data should be available through:
    - S3
    - NFS
    - Web portal
    - Share (equivalent to Amazon Simple Cloud Storage)
    - Samba
5. Selectable basic storage block size (data chunk)
6. Scalability:
    - All nodes in the cluster participate in data processing, ensuring performance proportional to the number of nodes involved
    - Hardware resources need not be homogeneous
7. Authorization and authentication
    - Ability to integrate with existing identity management systems such as LDAP, AD, Linux PAM, token-based S3 and SSO (SAML 2.0, OpenID Connector)
    - The solution should also provide the ability to use a REST API to manage resources.
    - Ability to delegate privilege management.
8. Accounting and control of resource consumption
    - Creating detailed reports on resource usage by users
    - Create limits on maximum user disk space usage (quota) and limits on data transfer rates and storage time limits.
9. The solution should be able to use AES-256 data encryption (inflight and at rest)
10. The solution must proactively check the integrity of the data and fix any problems that are detected.
11. Management via API, console (CLI) and web.
12. Perpetual license to use existing cluster resources.

2.5.3 Network architecture

The cloud will be launched on the existing network infrastructure (1G or 40G connections), at the same time the cloud software must enable cooperation with network infrastructure of higher throughput (100G).

The network infrastructure of the A-CHOICeM cloud must be implemented in such a way as to ensure efficient and reliable connections forming the access network for users and the entire network for cloud management and access to storage resources.

The network connection requirements have been grouped as "DATA network" and "MGMT network".

DATA network, it is a network designed to:

1. access to/from the Internet for cloud resources
2. access to/from UW ICM network for cloud resources
3. access to storage resources DATA network must allow:
   - connection to a single cloud node with throughput of at least 40G
   - connection to the resources of the PIONIER network and the Internet with a capacity of at least 100G
   - Creating virtual networks, assigning addresses from private and public address pools, dynamically and statically assigned.
   - Connections to/from the Internet made by means of assigned public IP addresses.
   - Connections to the Internet realized as SNAT connections with the use of shared public IP addresses
   - connections from the Internet realized as DNAT connections using shared public IP addresses
   - accountability of resources assigned to users (in particular the use of shared IP addresses)
   - Network scaling (increasing IP addressing resources by additional pools) MGMT network is a network designed to: connections to manage, monitor and configure cloud resources

MGMT network must enable connections between cloud nodes with at least 1G bandwidth

In addition, the architecture must meet the following requirements:
- DNAT and SNAT functionality must be implemented within the virtual resources of the cloud nodes.
- The DNAT function must be configured by the user within the allocated limits.
- The Contractor is obliged to reconfigure the Cloud once if the Ordering Party upgrades the Cloud network infrastructure during the guarantee period.
- The solution should be able to define SDN (Neutron SDN), compute nodes should have direct access to part of the network (L2) and indirect access to external networks through dedicated network nodes.
- Cloud network nodes must provide redundancy.

2.5.4 Cloud architecture and services

1. The cloud will be based on **OpenStack** environment. The entire system should be built using Open Source components to the greatest extent possible. In the case of using components with licenses other than Open Source, the supplied license should allow for free updates (especially security) within 3 years from the moment of delivery, and allow for lifetime operation also in the case of lack of support.
2. OpenStack controller servers (controller nodes) will be placed on a virtualization cluster (undercloud) built with the use of HA technology, allowing for seamless migration of servers, adding and removing servers from the cluster, backup and recovery of controllers, in a manner that does not disrupt the operation of the cloud solution.
3. The cloud deployment mechanism should be reproducible (the delivered installation "scenario" must allow the initial state of the cloud to be recreated from scratch - i.e. from bare-metal to ready-for-use).
4. On a cluster of controllers, space should be provided for virtual machines supporting associated services (portals, template repositories and configuration, IdM, monitoring and logging, etc.), the size of this cluster should be planned accordingly.
5. The cloud environment being built should provide a minimum set of services consisting of:
   - Adding/removing/managing servers to cloud infrastructure (metal as a service).
   - Services to manage instances, network and user and project access on an IaaS platform
   - Storag components (type: image, volume, S3, etc.) for virtual machines, physical machines and containers. Support for multi-tier eg: S3 built on disks of different types, locations or availability.

- Networking service compliant with OpenStack Networking API - providing such things as management of networks, subnets, routers, load balancers, VPNs, firewalls, IP addressing.
- Central client authentication, service discovery, and distributed access control service compliant with OpenStack Identity API.
- Service to collect and monitor system usage information. The service must collect information from OpenStack and other cloud components, allow monitoring and accounting of resource usage:
  - for billing purposes
  - for administrative purposes

7. Central system log collection service.
   - A service for managing quotas (limits on allocated resources) in a way that allows actions to be taken automatically when the indicated values are reached.
   - Cloud-based application orchestration service.
   - Resource Reservation Service.

All services should be constructed redundantly, ensuring high availability.

The cloud will be deployed using the existing storage infrastructure described in the !"#$ &$ to the POZ.

In addition, due to the functional requirements specified in this PO, it may be necessary to supply, install and configure other components in addition to those listed above. The Contractor must foresee and execute deliveries, installations and configurations of all necessary additional components within the scope of the order. These activities must be indicated and described in the offer.

2.5.5  Integration

All portals creating the cloud solution should be integrated using Single Sign-On (SSO) system implemented by the contractor, offering simultaneous support for different authentication mechanisms, min. PIONIER.Id, CAS UW, LDAP ICM (Keycloak) and one-time password sent to an email address. The last mechanism must be delivered during the implementation of the SSO system and must allow editing the list of allowed mail domains to which the password will be sent.

2.5.5.1 Integration with the Ordering Party's resources and services:

The Portal must allow access to resources managed by SLURM Workload Manager (Poplar, Okeanos systems). Access parameters will be requested by the user along with batch data through a generic template available in the Portal. The portal should allow the creation, modification and management of custom templates, including features that allow easy customization to the requirements of running applications - e.g. the ability to run multiple program runs on different batch data. After defining a task, the system should automatically order the task to be executed under the control of the queuing system. The mechanisms of working with applications (models of their use) are described in section 3. The portal should be designed to allow the execution of applications in batch mode (see section 3.1) and interactive mode (see section 3.2). Applications running in different workflows should be able to share disk space for further processing (pre/post processing). The solution should use disk resources that are part of the existing infrastructure (s3 storage). The user should be able to track running jobs, their status, parameters, preview generated output data and control the execution of jobs directly from the Portal to the extent offered by the SLURM system control programs such as 'scontrol' and others. Advanced use of these tools should be available through the ability for the user to enter their own instructions that will be redirected to the queue system.

Cloud storage integration required

1. The solution resides on existing hardware, the contractor will have one full Enigma module to use.
2. Storage cloud should be integrated with other elements of the cloud solution (same identities, authentication mechanism, mechanism of granting privileges)
3. Desired integration of existing S3 (Quantum) and Lustre solution to remove walls between separate resources data storage (Quantum/Enigma//Tetyda), e.g. using data providers like OneData/DataCore(ex-Caringo)
4. The computational part can use the data from the storage part and vice versa (e.g.: it is possible to extract data from the storage part and then, after generating a result, you can put this result back into the storage part)
5. The part responsible for "metering/billing" should be integrated and include both the computing and the cloud part.

5.4.2 Integration with external resources:

The user portal should allow access to quantum computing I AI and ML portals, bioinformatics, and such as Garuda, Taxila, graph computing (Urica XC, Trovares, Jena).

The portal must allow the user to move, at will, to popular public clouds Azure, AWS, IBM, Google and others.

## 2.6 Documentation

The bidder will provide documentation including:
1. Full editable documentation of the implementation in the scope of describing the configuration of servers, networks, storage, services and software. In particular, the documentation should include information about the network topology, addressing, connection schemes, configuration files of the deployed services and configured devices, versions of the used software, passwords.
2. A description of the components used in the solution and their interrelationships and relationships.
3. The procedure used to implement the cloud.
4. A description of the basic procedures for using the cloud. In this respect, at a minimum, the following issues should be covered:
   - Procedure for enabling and disabling a cloud cluster (both undercloud and overcloud)
   - Backup and recovery procedures
   - Failover procedures - restoring services after a failure of any server or service
   - Fallback procedures - description of actions in case of power supply failure or hardware failure of an infrastructure element
   - Cloud Software Update Procedure
   - Portal reinstallation procedure
   - Procedure for updating SSL certificates
   - Procedure of expanding the cluster with new servers or switches
   - Cloud operations from an administrator's point of view e.g. adding a new identity provider, changing service policies, adding
   - new networks, setting quotas, establishing user hierarchies and assigning permissions to the local administrator, etc.
   - Instructions for creating and adding new applications and service templates to the catalog. Examples of ready-made and documented templates for appointing virtual machines, kubernetes clusters and applications.
   - Description of the monitoring system and the central operation logging service, as well as description of basic operations related to these services (configuration and usage, e.g. architecture, backups, information extraction, etc.)
   - Procedure to deal with an attack from an external/internal IP address
   - Procedure to be followed in the event of a break-in to an account (authorized user, administrator)
   - Billing (REST API description, data format, export, import).
5. API description
6. A practical cloud user's guide. Instructions on how to perform basic operations in the cloud, addressed not necessarily to technical people (e.g. how to start a virtual machine or an application / service).
7. Diagram and description of how future possible expansion will take place.

## 2.7 Training

Within 3 months of contract acceptance, the contractor will provide training to the contracting authority's administrators:

- Cloud deployment - discussion of tools and components used, assumptions made, method used, localization of installation scripts and adjusting them to the needs of the ordering party.
- Current maintenance, basic operations, software upgrades, emergency shutdown, uplift, restore from backup, hardware replacement, expansion, network reconfiguration.
- Detailed discussion of network architecture (tunneling, routing) in the system and end user parts. Mutual interactions between particular components (bridge, protocols)
- Installation and updating of user applications, images, privilege management

The bid will include a training schedule.

## 2.8 Warranty and support

1. All technical devices supplied under the contract will be covered by at least a 24-month manufacturer's warranty, provided at the place of installation, ensuring that in case of failure the device is replaced with an operational one within 48 hours of notification.
2. Furthermore, the Contractor shall provide the Ordering Party with a warranty for the delivered software for a period of 24 months from the date of order acceptance. During the warranty period, the Contractor shall ensure that the software operates in accordance with these OPZ and the supplied documentation. Licenses for the provided software are perpetual. In case of using any paid licenses or licenses with limited validity period, the cost of these licenses in the following 48 months shall be presented in the offer.
3. The Contractor shall ensure the possibility of extending the warranty and support at a price not higher than specified in the offer for additional years for a period of at least 48 months from the date of acceptance of the order.

4. During the warranty period, the software will be covered by the contractor's support guaranteeing

- Response time to report no longer than 4 hours
- Time to fix critical error not longer than 48h
- Time to fix the error not critically longer than 10 working days
- In addition, a specified, not less than 40 hours of specialist consultation per year for troubleshooting and software updates.
- Moreover, during the support period, the contractor will perform a one-time reconfiguration of the network, free of charge, if the ordering party expands the network during this period (from 1GE to 40GE). This reconfiguration will be performed on an agreed date indicated by the ordering party at least 14 days in advance.
- The contractor will support administrators in upgrading the OpenStack release to the next "Extended Maintenance" version.
- The Contractor undertakes to support administrators in the expansion of the cloud with resources released from the Enigma module (in cooperation with administrators, the Contractor will expand the existing solution with new resources (servers, switches)).

## 2.9 Implementation and payment schedule

The functionalities described in points 1-8 of the OPZ will be implemented within a period not longer than 180 days from the conclusion of the agreement.

The trainings will be implemented according to the training schedule in a period not exceeding 180 days from the conclusion of the agreement.

## 2.10 Way of collecting the order

The execution of the order shall be confirmed by acceptance tests as specified in the Annex to the contract.

===============================================

ACKNOWLEDGMENT

The ICM team involved in this work consisted of: Marek Michalewicz (project initiator and general project manager), Wojciech Sylwestrzak (project management), Marzanna Zieleniewska (public procurement), Jarosław Skomiał (network and cloud architecture), Grzegorz Bakalarski, Rafał Maszkowski, Mirosław Nazaruk, Robert Paciorek, Marcin Semeniuk, Sebastian Tymkow (hardware and storage), Michał Dzikowski, Grzegorz Gruszczyński, Michał Hermanowicz, Maciej Szpindler, (software and cloud architecture) Joanna Jędraszczyk (user aspects), Krzysztof Młynarski (cybersecurity), Robert Sot (University procedures, regulations and relationship). These colleagues contributed to the technical specifications of this project. The author believes that difficult and negative experiences in research and academic life should also be reported for the benefit of openness, honesty, integrity and learning.

The author wishes to express his deep appreciation and gratitude for the involvement in this project and for individual contributions of all the above-mentioned colleagues.

REFERENCES

1. OPENAire National Initiatives – Poland, https://www.openaire.eu/os-poland
2. Armbrust, M., et al. (2010). A view of cloud computing. Communications of the ACM, 53(4), 50-58.
3. Buyya, R., et al. (2009). Cloud computing and emerging IT platforms: Vision, hype, and reality for delivering computing as the 5th utility. Future Generation Computer Systems, 25(6), 599-616.
4. Gonzalez et al. „Cloud resource management: towards efficient execution of large-scale scientific applications and workflows on complex infrastructures", Journal of Cloud Computing: Advances, Systems and Applications (2017) 6:13
5. Beckman, P. H., "Building the TeraGrid", (2005) Philosophical Transactions of The Royal Society A Mathematical Physical and Engineering Sciences 363(1833):1715-28, DOI: 10.1098/rsta.2005.1602
6. Open Science Grid. (2021). About OSG. Retrieved from https://www.opensciencegrid.org/, last accessed 13 April 2023
7. XSEDE USA, https://web.archive.org/web/20220820025648/https://portal.xsede.org/#/guest, last accessed 16 April 2023
8. J. Towns et al., "XSEDE: Accelerating Scientific Discovery," in Computing in Science & Engineering, vol. 16, no. 5, pp. 62-74, Sept.-Oct. 2014, doi: 10.1109/MCSE.2014.80
9. Advanced Cyberinfrastructure Coordination Ecosystem, https://access-ci.org/ last accessed 16 April 2023
10. PL-Grid – Poland, https://www.plgrid.pl , last accessed 16 April 2023
11. R. Buyya, D. Abramson and J. Giddy, "Nimrod/G: an architecture for a resource management and scheduling system in a global computational grid," Proceedings Fourth International Conference/Exhibition on High Performance Computing in the Asia-Pacific Region, Beijing, China, 2000, pp. 283-289 vol.1, doi: 10.1109/HPC.2000.846563
12. European Open Science Cloud (EOSC) - An initiative of the European Commission to build a trusted and open virtual environment for the sharing and reuse of research data and services. (https://eosc.eu/), last accessed 13 April 2023
13. Australian Research Data Commons (ARDC) - An Australian initiative to enable the sharing, publishing, and reuse of research data and services through the development of an open science cloud infrastructure. (https://ardc.edu.au/), last accessed 13 April 2023
14. National Science Foundation's CloudBank - A US initiative that provides cloud computing resources and services to researchers funded by the NSF, with the goal of making cloud computing more accessible to researchers. (https://www.cloudbank.org/), last accessed 13 April 2023
15. National Institutes of Health's Science and Technology Research Infrastructure for Discovery, Experimentation, and Sustainability (STRIDES) Initiative - A US initiative that provides researchers with access to cloud-based computing and data resources to accelerate biomedical research. (https://datascience.nih.gov/strides), last accessed 13 April 2023
16. Nordic e-Infrastructure Collaboration (NeIC) - An initiative of Nordic countries to develop and operate e-Infrastructure services for researchers across the region, including cloud computing resources. (https://neic.no/), last accessed 13 April 2023
17. Science and Technology Facilities Council (STFC) Cloud Portal - A UK initiative that provides researchers with access to a range of cloud computing resources, including public cloud providers and private cloud environments. https://dareuk.org.uk/ , last accessed 13 April 2023
18. Helix Nebula - The Science Cloud - A European initiative that aims to establish a federated cloud infrastructure to provide European

researchers with access to computing and data resources. (https://www.helix-nebula.eu/), last accessed 13 April 2023
19. Australian Research Data Commons (ARDC) Nectar Research Cloud, https://ardc.edu.au/services/ardc-nectar-research-cloud/ , last accessed 16 April 2023
20. Ubercloud   https://www.theubercloud.com/, last accessed 13 April 2023
21. https://www.theubercloud.com/ubercloud-compendiums, last accessed 13 April 2023
22. Rescale https://rescale.com/  https://rescale.com/cloud-hpc/, last accessed 13 April 2023
23. XTREME-D  https://xtreme-d.net/, last accessed 13 April 2023
24. XTREME Stargate The New Era of HPC Cloud Platforms https://www.youtube.com/watch?v=QdfHnn9L7YY&ab_channel=hpcaiadvisorycouncil, last accessed 13 April 2023
25. Ronin   https://ronin.cloud/, last accessed 13 April 2023
26. Chameleon: A configurable experimental environment for large-scale edge to cloud research, https://www.chameleoncloud.org/ https://scholar.google.com/citations?user=A4pT-T4AAAAJ&hl=en   last accessed 16 April 2023
27. https://www.bbc.com/historyofthebbc/anniversaries/february/yes-minister/ last accessed 16 April 2023
28. Academic Computer Centre Cyfronet AGH https://www.cyfronet.pl/en/4421,main.html last accessed 16 April 2023
29. SZCZEGÓŁOWY OPIS OSI PRIORYTETOWYCH PROGRAMU OPERACYJNEGO INTELIGENTNY ROZWÓJ 2014-2020 https://www.funduszeeuropejskie.gov.pl/media/90780/SZOOP_aktualizacja_czerwiec_2020.pdf p. 6,
30. Supercomputing Frontiers Europe 2020 and 2021 https://supercomputingfrontiers.eu/2020/, last accessed 13 April 2023 https://supercomputingfrontiers.eu/2021/conference-programme/, last accessed 13 April 2023
31. NICOLAS TONELLO, "Constellation® – Supercomputing at your fingertips – Delivering HPC power and expertise to all" https://www.youtube.com/watch?v=j1tUJ6PaNf4&t=2s&ab_channel=ICMUniversityofWarsaw, last accessed 13 April 2023
32. K. Kahey, "Chameleon: Taking Science from Cloud to Edge", https://www.youtube.com/watch?v=ekBAZ1-t9Qk&t=43s&ab_channel=ICMUniversityofWarsaw, last accessed 13 April 2023
33. Wolfgang Gentzsch, "Using distributed HPC technology for building an automated, self-service, truly multi-cloud simulation platform", https://www.youtube.com/watch?v=8uFU59xO6xo&t=18s&ab_channel=ICMUniversityofWarsaw, last accessed 13 April 2023
34. Edward Hsu, "High Performance Computing Built For Cloud Using an Intelligent Control Plane Approach" https://www.youtube.com/watch?v=JeOsejMV2mg&t=44s&ab_channel=ICMUniversityofWarsaw, last accessed 13 April 2023
35. Tara Madhyastha, "RONIN: Secure Self-service High Performance Research Computing in the Cloud", https://www.youtube.com/watch?v=EmMEd4W0AmU&t=19s&ab_channel=ICMUniversityofWarsaw, last accessed 13 April 2023
36. Valerie E. Polichar, "Creating a Technology Vision: Planning for the Next Generation of Research Support", https://www.youtube.com/watch?v=osWXOVCNVgU&t=1s&ab_channel=ICMUniversityofWarsaw
37. Blueprint for the Digital University. A Report of the UCSD Research Cyberinfrastructure Design Team (4/24/09), https://research-it.ucsd.edu/otherresources/UCSD-Blueprint-for-the-Digital-University.pdf
38. Minor, D., & Polichar, V. E. (2022). Blueprint 2030: The Digital University in the Next Decade. *UC San Diego: Library*. Retrieved from https://escholarship.org/uc/item/920165p2
39. www.openfoam.com  last accessed 16 April 2023
40. https://www.quantum-inspire/  last accessed 16 April 2023

**Marek Michalewicz** Marek Michalewicz obtained PhD in Theoretical Physics at the ANU Canberra in 1987. He is an adviser and a member of Technical Executive Committee at the National Supercomputer Centre in Singapore, a founder of Siranga start-up building quantum tunnelling-based photodetectors. He founded and run four other start-ups previously.
He was the Director of the Interdisciplinary Centre for Mathematical and Computational Modelling, University of Warsaw from July 2018 till the end of 2021, CEO of A*STAR Computational Resource Centre in Singapore (2009-2016) and a Principal Research Scientist in High Performance Computing and Communication Group, Division of Mathematical Sciences, in CSIRO, Australia (1990-2000). He was a co-inventor and the leader of InfiniCortex project (2014-2016) to build a global scale concurrent, InfiniBand (RDMA) connected supercomputer located in seven countries on four continents. He was also one of the main architects and driving forces behind creation of NSCC in Singapore. Marek does research in Supercomputing, Computational Science, Theoretical Physics and NEMS sensors.